\documentclass[aps, prd, twocolumn, superscriptaddress]{revtex4-1}
\usepackage{graphicx}
\usepackage{amssymb}
\usepackage{amsmath}
\usepackage{booktabs}
\usepackage[colorlinks=True,linkcolor=blue,anchorcolor=blue,citecolor=blue,CJKbookmarks=true]{hyperref}

\usepackage{ulem}

\begin{document}
\title{Backreaction effect and plasma oscillation in pair production for rapidly oscillating electric fields}

\author{R. Z. Jiang}
\affiliation{School of Science, China University of Mining and Technology, Beijing 100083, China}
\affiliation{State Key Laboratory for GeoMechanics and Deep Underground Engineering, China University of Mining and Technology, Beijing 100083, China}

\author{C. Gong}
\affiliation{North China Electric Power University, Baoding 071003, China}

\author{Z. L. Li}
\email{Corresponding author. zlli@cumtb.edu.cn}
\affiliation{School of Science, China University of Mining and Technology, Beijing 100083, China}
\affiliation{State Key Laboratory for GeoMechanics and Deep Underground Engineering, China University of Mining and Technology, Beijing 100083, China}

\author{Y. J. Li}
\email{Corresponding author. lyj@aphy.iphy.ac.cn}
\affiliation{School of Science, China University of Mining and Technology, Beijing 100083, China}
\affiliation{State Key Laboratory for GeoMechanics and Deep Underground Engineering, China University of Mining and Technology, Beijing 100083, China}

\date{\today}

\begin{abstract}
The backreaction effect and plasma oscillation in pair production for rapidly oscillating electric fields are investigated by solving quantum Vlasov equation.
Contrary to previously thought, it is found that the backreaction effect can be neglected in the pair production for a rapidly oscillating but weak electric field, particularly, for a subcritical external electric field with frequency chirp.
In some cases the oscillation period of created electron-positron plasma can be described by a simple formula constructed based on the Langmiur oscillation frequency, but it is impossible in general case because the plasma oscillation period directly depend not only on the final number density of created particles, but also on the external electric field parameters.
Moreover, it is found that the momentum spectrum presents complex interferences after considering the backreaction.
These results give us the safety range of external electric fields for taking no account of the backreaction effect and deepen our understanding of the pair production with the backreaction effect.
\end{abstract}

\maketitle

\section{INTRODUCTION}

Since Dirac proposed the relativistic wave equation and predicted the existence of positrons \cite{Dirac1928}, many researches have been done on how to produce electron-positron pairs from vacuum.
Sauter \cite{Sauter1931} found that the electron-positron pairs can be produced from vacuum in a strong static electric field by tunneling mechanism.
And then, Schwinger \cite{Schwinger1951} calculated the pair production rate in a constant electric field by the proper-time method and gave out the critical electric field strength $E_{\rm{cr}}=m^2/e\approx 1.3\times10^{18}\rm{V}/\rm{m}$ (the natural units $\hbar=c=1$ are used).
As a result of these pioneering works, the phenomenon of electron-positron pairs produced from the vacuum in intense external fields is also known as the Sauter-Schwinger effect \cite{Xie2017,Fedotov2022}.
The laser intensity corresponding to the critical electric field strength is $I\approx 10^{29}\rm{W}/\rm{cm}^2$, which is much larger than the possibility of the current laser facilities \cite{Tanaka2019, Tiwari2019, Yoon2021}, so it is not yet possible to produce observable electron pairs in experiments.
However, the already operating X-ray free electron laser (XFEL) at DESY in Hamburg is expected to achieve subcritical fields of $E\approx 0.01-0.1E_{\rm{cr}}$ \cite{Ringwald2001}.

To reduce the threshold of pair production, several approaches have been proposed to produce observable electron pairs under subcritical electric field conditions.
One method that can effectively enhance the pair production is the dynamically assisted Schwinger mechanism \cite{Schutzhold2008,Nuriman2012,Fey2012,Kohlfurst2013,Li2014}, in which the electric field adopts a combination of a low-frequency strong field and a high-frequency weak field.
Another approach is to use frequency chirp to increase the pair yield by increasing the effective frequency of the electric field.
In Refs. \cite{Jiang2013,Abdukerim2017,Gong2020,Olugh2019,Wang2021}, authors studied the effect of spatially homogeneous electric fields with frequency chirp on the pair production by the quantum kinetic equation and found that for some chirp parameters the pair yield could be improved three to four orders of magnitude.
Moreover, the enhancement effect of pair production in a spatially inhomogeneous electric field  with the frequency chirp were also found \cite{Li2021,Mohamedsedik2021}.

Previous studies have shown that the electrons and positrons produced from vacuum can be accelerated in an external field and form a new electric field, which is called the internal electric field \cite{Alkofer2001,Bloch1999,Roberts2002,Pla2021}.
The internal electric field can affect the generation of particle pairs together with the external electric field, that is the backreaction.
For subcritical external field strength, the number of produced particle pairs is relatively less, and the backreaction is generally considered to be negligible.
However, in the study of pair production in a frequency chirp electric field, if the field lasts for a long time, the effective frequency will be very large and may even exceed the primary frequency, which can greatly enhance the multi-photon pair production process and produce a large number of particle pairs.
Therefore, whether the backreaction effect in this case can be negligible is a problem worth researching. In this paper, we will figure out the parameter scope of external electric fields in which the backreaction effect cannot be ignored.

In addition, when the backreaction effect is considered a plasma oscillation will occur.
In Ref. \cite{Alkofer2001}, the plasma oscillation in pair production for a sinusoidal electric field is studied by quantum Vlasov equation (QVE) and a ultrarelativistic formula is given to estimate the oscillation period of the plasma.
The ultrarelativistic formula shows that the oscillation period only depends on the maxima of particle number density and has no direct relation with the field parameters.
In this work, we will study the plasma oscillation for a rapidly oscillating electric field by QVE, explore the determining factors of the oscillation period, and check whether this formula still holds.

The structure of this paper is as follows: Section \ref{sec:two} briefly introduce the theoretical framework of quantum dynamics-based backreaction effects.
 Section \ref{sec:three} is our numerical results: Subsection \ref{A} discusses the effect of the backreaction effect on the final particle number density; subsection \ref{B} gives the effect of the backreaction effect on the momentum spectra; subsection \ref{C} study the relation between the plasma oscillation period and the number density of produced particle pairs. Section \ref{sec:four} is a summary and discussion.

\section{THEORETICAL FORMALISM}
\label{sec:two}

The magnetic effects can be neglected when a standing-wave field produced by two counter-propagating laser pulses is considered.
The spatial scales for electron-positron pairs production is on the same order of magnitude as the Compton wavelength of the electron, which is far smaller than the focusing radius of the laser, so it can be assumed that the laser is spatially homogeneous.
By using the temporal gauge, $A_0= 0$, the spatially homogeneous and time-dependent four-vector potential can be written as $A_\mu= (0,0,0,A(t))$, and the external electric field is $E_{\rm{ext}}=-d A(t)/dt$. The form of the external electric field we used is
\begin{eqnarray}\label{eq:eext}
E_{\mathrm{ext}}\left( t \right) =E_0\mathrm{e}^{-\frac{t^2}{2\tau ^2}}\cos \!\:\left( \omega t \right) ,
\end{eqnarray}
where $E_0$ is the electric field amplitude, $\omega$ is the laser frequency, and $\tau$ is the pulse duration.

The key quantity to study the electron-positron pair production with the QVE is to obtain the momentum distribution function $f({\bf{k}},t)$.
For spatially homogeneous and time-dependent electric fields, ignoring collisions between particles, the distribution function is determined by $df({\bf{k}},t)/dt=S({\bf{k}},t)$, where $S({\bf{k}},t)$ denotes the source term of pair production.
When the external field strength is relatively large, the backreaction brought by the internal electric field is gradually reflected, so the total electric field $E_{\rm{tot}}(t)$ should be modified as the sum of the external field and the internal field $E_{\rm{int}}(t)$, i.e. $E_{\rm{tot}}(t)=E_{\rm{ext}}(t)+E_{\rm{int}}(t)$.
After considering the influence of the backreaction, we can get the coupled equations of the distribution function and the internal electric field
\begin{eqnarray}\label{eq:f}
\dot{f}(\mathbf{k},t)=\frac{eE_{\rm{tot}}(t)\varepsilon _{\perp}}{2\Omega ^2(\mathbf{k},t)}\int_{-\infty}^t&&\hspace{-3mm}dt'\frac{eE_{\rm{tot}}(t)\varepsilon _{\perp}}{\Omega ^2(\mathbf{k},t')}[1-2f(\mathbf{k},t')] \nonumber\\
&&\hspace{-3mm}\times \cos \!\:[2\int_{t'}^t{dt''\Omega (\mathbf{k},t'')}],
\end{eqnarray}
\begin{eqnarray}\label{eq:eintdot}
{{\dot E}_{{\rm{int}}}}\left( t \right) =  - 4e\int &&\hspace{-1mm}\frac{{{d^3}k}}{{{{\left( {2\pi } \right)}^3}}}\Big[\frac{{{k_\parallel }\left( t \right)}}{{{\rm{\Omega }}\left( {{\bf{k}},t} \right)}}f\left( {{\bf{k}},t} \right)\nonumber\\
&&\hspace{-3mm}+ \frac{{{\rm{\Omega }}\left( {{\bf{k}},t} \right)}}{{e{E_{{\rm{tot}}}}\left( t \right)}}\dot{f}\left(\bf{k},t\right)- \frac{{e{{\dot E}_{{\rm{tot}}}}\left( t \right)\varepsilon _ \perp ^2}}{{8{{\rm{\Omega }}^5}\left( {{\bf{k}},t} \right)}}\Big],
\end{eqnarray}
where the dot on the letter represents the first order time derivative, $\left|e\right|$ is the renormalized charge of the electron \cite{Bloch1999}, ${\bf{k}}=\left( {{{\bf{k}}_\bot },{k_\parallel}}\right)$ is the canonical momentum, ${k_\parallel }\left(t \right)={k_\parallel}-eA\left(t\right)$ is defined as the kinetic momentum along the external electric field, ${\varepsilon_\bot}=\sqrt{{m^2}+{\bf{k}}_\bot^2}$ is the transverse energy squared, $m$ is the mass of the electron, and ${\rm{\Omega }}\left({{\bf{k}},t}\right)=\sqrt{\varepsilon_\bot^2+k_\parallel^2\left(t\right)}$ is the total energy squared.
The quantum statistics effect and the non-Markov effect on the pair production can be seen from $[{1-2f({{\bf{k}},t'})}]$ in Eq. (\ref{eq:f}).
The first term on the right hand side of Eq. (\ref{eq:eintdot}) represents the conduction current, the second term is the polarization current, and the third term is the charge renormalization part added to eliminate the divergence of the polarization current.

For the convenience of numerical calculation, two auxiliary variables $u\left( {{\bf{k}},t} \right)$ and $v\left( {{\bf{k}},t} \right)$ are introduced
\begin{eqnarray}\label{eq:uv}
\begin{array}{l}
u\left( \mathbf{k},t \right) =\int_{t_0}^t{dt'W\left( \mathbf{k},t' \right) \left[ 1-2f\left( \mathbf{k},t' \right) \right]}\\
\\
\quad \quad \quad \quad     \times   \cos \left[ 2\int_{t'}^t{dt''\Omega \left( \mathbf{k},t'' \right)} \right] ,
\\
\\
v\left( \mathbf{k},t \right) =\int_{t_0}^t{dt'W\left( \mathbf{k},t' \right) \left[ 1-2f\left( \mathbf{k},t' \right) \right]}\\
\\
\quad \quad \quad \quad      \times  \sin \left[ 2\int_{t'}^t{dt''\Omega \left( \mathbf{k},t'' \right)} \right] ,
\end{array}
\end{eqnarray}
then equation (\ref{eq:f}) can be equivalently transformed into the following first-order differential equations
\begin{eqnarray}\label{eq:fuv}
\begin{array}{l}
\dot{f}\left( {{\bf{k}},t} \right)= \frac{1}{2}W\left( {{\bf{k}},t} \right)u\left( {{\bf{k}},t} \right),\\
\\
\dot{u}\left( {{\bf{k}},t} \right)= W\left( {{\bf{k}},t} \right)\left[ {1 - 2f\left( {{\bf{k}},t} \right)} \right] - 2{\rm{\Omega }}\left( {{\bf{k}},t} \right)v\left( {{\bf{k}},t} \right),\\
\\
\dot{v}\left( {{\bf{k}},t} \right)= 2{\rm{\Omega }}\left( {{\bf{k}},t} \right)u\left( {{\bf{k}},t} \right),
\end{array}
\end{eqnarray}
where $W\left( {{\bf{k}},t} \right) = e{E_{{\rm{tot}}}}\left( t \right){\varepsilon _ \bot }/{{\rm{\Omega }}^2}\left( {{\bf{k}},t} \right)$.

It can be seen from Eq. (\ref{eq:eintdot}) that to obtain the internal electric field at time $t$, the momentum distribution function at the same time must be integrated, but the internal electric field at the same time must be known to calculate the momentum distribution function at time $t$.
To solve this contradiction, we calculate the internal electric field by iteration.
First, we use the internal electric field at the previous time $(t-\Delta t)$ to calculate the total electric field at time $t$ by
\begin{equation}\label{eq:etot}
E_{\rm{tot}}(t) = E_{\rm{ext}}(t) + E_{\rm{int}}^l(t),
\end{equation}
where $l= 1,2, \cdots $ is the number of iterations.
Note that for the first iteration ($l=1$), the internal electric field $E_{\rm{int}}^1(t) = E_{\rm{int}}(t-\Delta t)$.
Using Eq. (\ref{eq:etot}), the momentum distribution function at time $t$ can be obtained by solving Eq. (\ref{eq:fuv}).
Then the internal electric field at time $t$ can be solved by Eq. (\ref{eq:eintdot}).
This is the first iteration.
Replacing the internal electric field in Eq. (\ref{eq:etot}) with the new one, the second iteration begins.
When the internal electric field satisfies our preset control condition $|E_{\rm{int}}^{l+ 1}(t)-E_{\rm{int}}^l(t)|<\rm{\varepsilon}$, where $\rm{\varepsilon}$ is a very small number, it can be considered that the real internal electric field at time $t$ has been obtained.
Plugging it into the total electric field and solving Eq. (\ref{eq:fuv}), the momentum distribution function at time $t$ can be solved as well.

The initial state of the system is a vacuum state without particles, so the single particle distribution function and auxiliary function satisfy the initial conditions $f\left( {{\bf{k}}, - \infty } \right) = u\left( {{\bf{k}}, - \infty } \right) = v\left( {{\bf{k}}, - \infty } \right) = 0$.
The initial condition of the internal electric field is ${E_{{\rm{int}}}}\left( { - \infty } \right) = 0$.
After obtaining the single-particle distribution function, integrating it in the full momentum space can obtain the resulting particle number density

\begin{eqnarray}\label{eq:n}
n\left( t \right) =2\int{\frac{d^3k}{\left( 2\pi \right) ^3}f\left( \mathbf{k},t \right)}.
\end{eqnarray}
The coefficient $2$ comes from the spin degeneracy of the fermions.

\section{NUMERICAL RESULTS AND ANALYSIS}
\label{sec:three}

\subsection{Particle number density}\label{A}

\begin{figure}[!ht]
\centering
\includegraphics[width=0.45\textwidth]{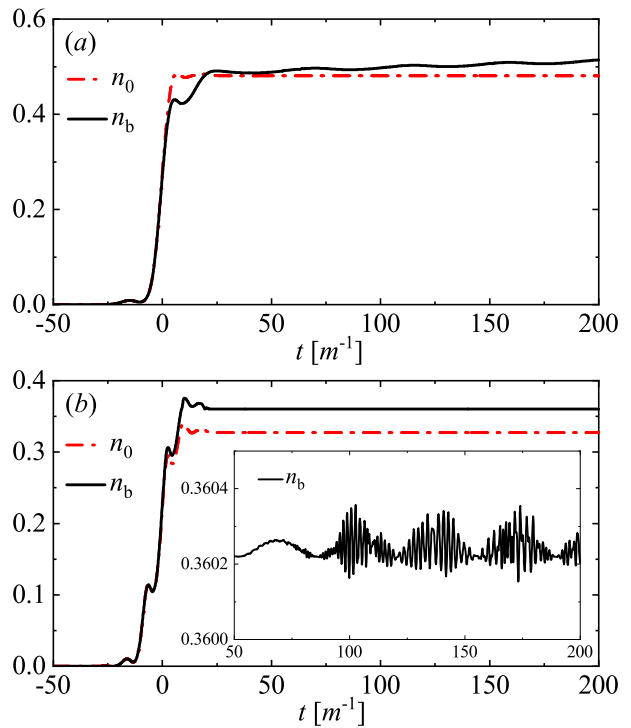}%
\caption{The particle number density as a function of time with (solid black lines) and without backreaction (dashed red lines). The external electric field parameters in (a) are $E_0=4.0E_{\rm{cr}}$, $\omega=0.15m$ and $\tau=12.0/m$, and in (b) are $E_0=4.0E_{\rm{cr}}$, $\omega=0.35m$ and $\tau=12.0/m$.
\label{fig:number1}}
\end{figure}

After considering the backreaction effect, when the external electric field vanishes the total electric field is not zero due to the existence of the internal electric field and the particle number density still changes with time.
So it is not easy to obtain a definite particle number density.
However, in a wide range of external field parameters, the oscillation of particle number density induced by the internal electric field is very very small, for example the second case we considered below, it is reasonable to define this relatively stable number density as the real particle number density.
The detailed explanation is as follows.
In Fig. \ref{fig:number1}, we show the number density of created pairs $n_b$ in the presence and $n_0$ in the absence of backreaction for two supercritical electric fields.
The field frequencies in Fig. \ref{fig:number1} (a) and (b) are $0.15m$ and $0.35m$, respectively.
Other field parameters are $E_0=4.0E_{\rm{cr}}$ and $\tau=12.0/m$.
It can be seen that the particle number density is invariable when the backreaction is not considered, but the situation is different when the backreaction is considered.
For such a supercritical external electric field, when the field frequency is relatively small, such as $\omega=0.15m$ in Fig. \ref{fig:number1}(a), the number density gradually increases with time, which indicates that the internal electric field induces the pair production.
However, when the frequency increases to a certain value, such as $\omega=0.35m$ in Fig. \ref{fig:number1}(b), the particle number density remains nearly constant because the internal electric field is not strong enough to stimulate sufficient particle pairs.
Enlarging the curve of $n_b$ in Fig. \ref{fig:number1}(b), one can see that the particle number density oscillates with time due to the existence of the internal electric field, but its variation range does not exceed $2.0\times10^{-4}m^{-3}$.
Therefore, the relatively stable number density can be considered as the real particle number density within the range of error allowed.
Our following calculation always meets this condition.

\begin{figure}[!ht]
\centering
\includegraphics[width=0.45\textwidth]{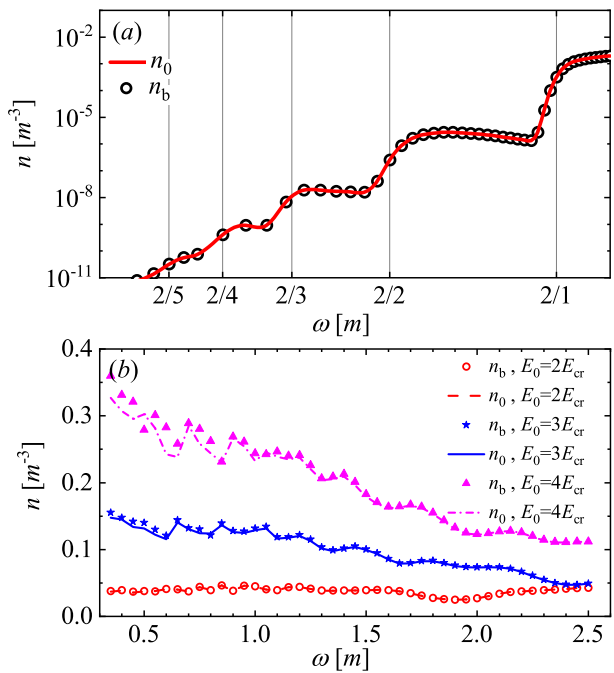}%
\caption{Particle number density changing with the field frequency for different field amplitudes. In (a), $E_0=0.1E_{\rm{cr}}$, and in (b), $E_0=2.0E_{\rm{cr}}$, $3.0E_{\rm{cr}}$, $4.0E_{\rm{cr}}$, the value of $\tau$ in both figures is $12.0/m$.
\label{fig:number2}}
\end{figure}

The particle number density varying with the field frequency for different external electric field amplitudes is shown in Fig. \ref{fig:number2}.
It can be seen that for the subcritical field with $E_0=0.1E_{\rm{cr}}$, see Fig. \ref{fig:number2}(a), the number density with and without the backreaction effect is almost the same, and the multi-photon absorption is obvious \cite{Kohlfurst2014}.
At the frequency $\omega=2m/{N_0}$ with the number of absorbed photons $N_0$, the particle number density increases greatly.
In the figure, $1-, 2-, 3-, 4-,$ and $5-$photon absorption are marked by vertical lines.
However, when $E_0=2.0E_{\rm{cr}}$, $3.0E_{\rm{cr}}$, and $4.0E_{\rm{cr}}$, see Fig. \ref{fig:number2}(b), the multiphoton pair production is not obvious, and simply increasing the frequency of the external field does not enhance the pair production but suppress it.
According to the Keldysh adiabatic parameter \cite{Keldysh} $\gamma=m\omega/e{E_0}$, we can know that in our calculation $\gamma  \sim O\left( 1 \right)$.
In this range, the tunneling pair production coexists with the multiphoton pair production, and their competitive relation is unfavourable for the pair production.
In addition, although the photon energy is high, the number density of produced particles is also affected by other field parameters.
When $E_0=2.0E_{\rm{cr}}, 3.0E_{\rm{cr}}$, the backreaction effect is still insignificant, but when $E_0=4.0E_{\rm{cr}}$, the backreaction effect of the internal electric field begins to affect the particle number density, see $\omega  \le 0.85m$.

\begin{figure}[!ht]
\centering
\includegraphics[width=0.45\textwidth]{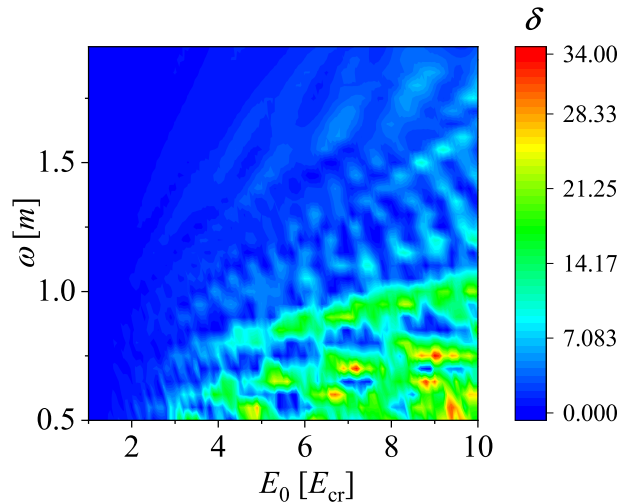}%
\caption{The difference degree of pair number density (DDOPND) $\delta$ as a function of the external field amplitude and frequency. The value of $\tau$ in this figure is $12.0/m$. In the calculation, the interval of $E_{0}$ is $0.1E_{\rm{cr}}$, and the interval of $\omega$ is $0.05m$.
\label{fig:error}}
\end{figure}

To further investigate the region where the backreaction effect cannot be ignored, we define the difference degree of pair number density (DDOPND) as $\delta=\left| {{n_b} - {n_0}} \right|/n_0\times 100\%$ and study its changes with the external field parameters $E_{0}$ and $\omega$, see Fig. \ref{fig:error}.
Although the supercritical field strength is used in the calculation, which is far away from the current experimental conditions, some interesting results can still be obtained.
In our results, for $E_{0}\le2.0E_{\rm{cr}}$, the maximum value of $\delta$ is about $1.93\%$ at $E_{0}=2.0E_{cr}, \omega=0.5m$ and the backreaction effect can be ignored.
For $2.0E_{\rm{cr}}<E_{0}\le3.0E_{\rm{cr}}$, the value of $\delta$ is generally within $5\%$, and in the region where $\omega$ is small, the DDOPND value may exceed $5\%$.
When the electric field amplitude $E_{0}$ is large and the frequency $\omega$ is small, i.e., the Keldysh parameter $\gamma$ is small, the DDOPND is large and the backreaction effect is more obvious.
This also shows that generally the backreaction effect can be neglected in the study of pair production in a high-frequency but weak electric field.
Therefore, it is safe to ignore the backreaction effect when study the pair production in a subcritical external field with frequency chirp.
This result is beyond our expectation.
In the very beginning, we thought that the particle number density would be improved greatly by the high-frequency photon absorption process and the backreaction effect became significant.
However, the actual result is not what we previously thought because of the competitive relation between the tunneling pair production and the multiphoton absorption.
In fact, the change of DDOPND with $E_0$ and $\omega$ is complex.
For example, when $E_{0}=7.0E_{\rm{cr}}$ and $\omega=0.8m$, $\delta\approx4.37\%$, the particle number density with and without backreaction are about $0.970m^{3}$ and $0.929m^{3}$, respectively.
The influence of the backreaction effect is almost unnecessary.
Whereas, when $E_{0}=7.0E_{\rm{cr}}$ and $\omega=0.9m$, $\delta\approx23.67\%$, the number density with and without backreaction are about $0.956m^{3}$ and $0.773m^{3}$, the backreaction effect is very obvious.

\subsection{Momentum spectrum}\label{B}

\begin{figure}[!ht]
\centering
\includegraphics[width=0.45\textwidth]{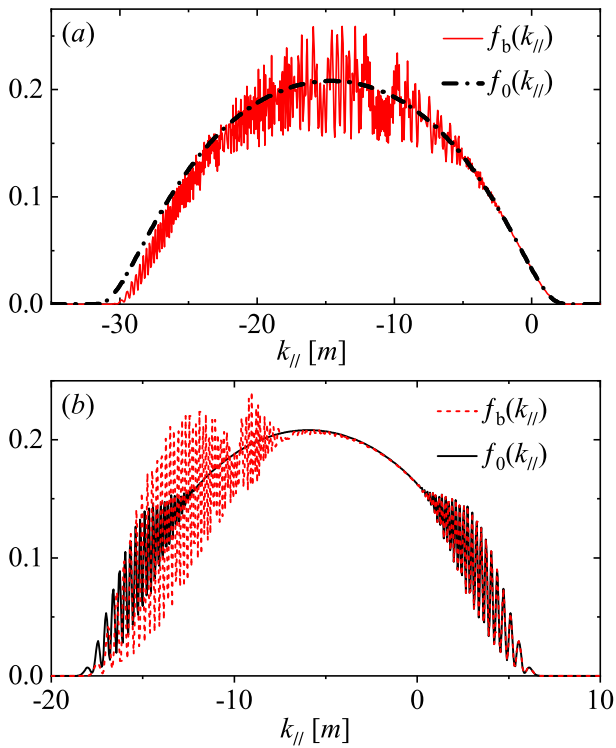}%
\caption{Comparison of the momentum distribution functions between with and without backreaction at $t=220.0/m$. The parameters of the external electric field in (a) are $E_0=2.0E_{\rm{cr}}$, $\omega=0.1m$, $\tau=12.0/m$, and in (b) are $E_0=2.0E_{\rm{cr}}$, $\omega=0.15m$, $\tau=12.0/m$.
\label{fig:distribution}}
\end{figure}

As shown in Fig. \ref{fig:distribution}, we compare the momentum distribution functions of created particles with and without backreaction, denoted by $f_{b}$ and $f_{0}$, respectively.
In Fig \ref{fig:distribution}(a), the momentum distribution $f_{0}$ is smooth because the particles are mainly generated by the main peak of the electric field, but the situation is different for slightly larger values of $\omega$.
As shown in Fig. \ref{fig:distribution}(b), although the sub-maximal peak of the electric field cannot produce sufficient particle pairs, the momentum distribution function $f_{0}$ shows obvious oscillations because of the infield interference.
Furthermore, without the backreaction, the momentum distribution functions $f_{0}$ is symmetric about the zero kinetic momentum in both cases.
However, with the backreaction, the momentum spectrum presents irregular oscillations and the symmetry is broken.
Another phenomenon is that the backreaction effect is observed in the left side of the momentum distribution functions while on the right side they are approximately identical to that without backreaction.
In fact, the change of the momentum spectrum with backreaction is very complex and sensitive to the parameters of the external electric field.

\subsection{Plasma oscillation period in pair production}\label{C}

\begin{figure}[!ht]
\centering
\includegraphics[width=0.45\textwidth]{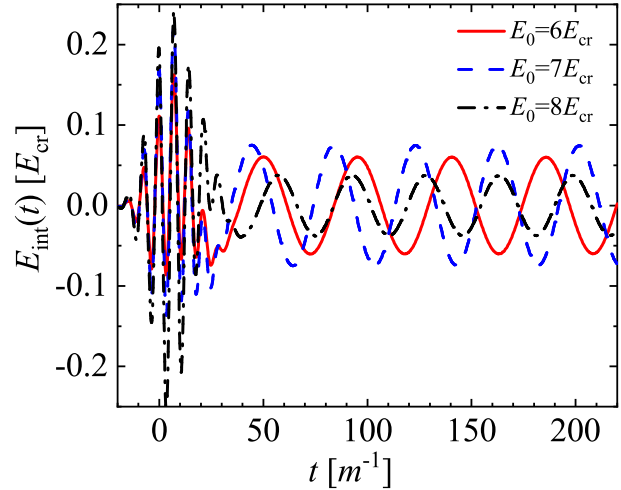}%
\caption{The internal electric field changes with time for $E_0=6.0E_{\rm{cr}}$, $7.0E_{\rm{cr}}$ and $8.0E_{\rm{cr}}$ respectively. Other field parameters are $\omega=0.9m$ and $\tau=12.0/m$.
\label{fig:eint}}
\end{figure}

We show the internal electric field evolution with time for different external field amplitudes in Fig. \ref{fig:eint}.
One can see when the external electric field exists ($t<40/m$), the amplitude of internal electric field increases with $E_{0}$, and its frequency is the same as that of the external electric field.
For instance, when $E_0=7E_{\rm{cr}}$, the peak value of the internal electric field is about $0.2E_{\rm{cr}}$.
The motion of particles is mainly dominated by the external electric field.
However, when the external electric field is turned off, the relation between the internal electric field and the external field parameters becomes complicated.
For example, the amplitude of the internal electric field does not increase monotonically with the $E_0$. Moreover, we already know that when the external electric field is strong enough, a large number of real electron-positron pairs will be generated, and these particle pairs will maintain a dynamic balance under the action of the internal electric field.
Therefore, the motion of created electron-positron pairs forms plasma oscillation.
It is worth noting that the period of plasma oscillation will have some relation with the number density of created particles and the parameters of the external electric field.
Previous studies have shown that the frequency of plasma oscillation decreases gradually with time and tends to a stable value at the end \cite{Benedetti2011}.
For the external electric field with high intensity and high frequency, the frequency of plasma oscillation can reach the stable state quickly.

Assuming that the external electric field disappears at $t_{0}$, then the internal electric field at $t>t_0$ can be approximately expressed as
\begin{eqnarray}\label{eq:eintfit}
{E_{{\rm{int}}}}\left( {t} \right) = {{{E}}_{{\rm{int}}0}}{\rm{cos}}\left( {\frac{{2\pi }}{{T}}t + \varphi } \right),
\end{eqnarray}
where $E_{\rm{int0}}$ is the amplitude of the internal electric field, $T$ is the period of oscillation, and $\varphi$ is the phase.
In this subsection, we mainly study the relationship between the oscillation period $T$ of the internal electric field and the number density of created particles $n$, the parameters of the external electric field $E_0$, $\omega$ and $\tau$.
In the following research, we study the relationship between $T$ and $n$ when $E_0$, $\omega$, and $\tau$ change, respectively.

\begin{figure}[!ht]
\centering
\includegraphics[width=0.48\textwidth]{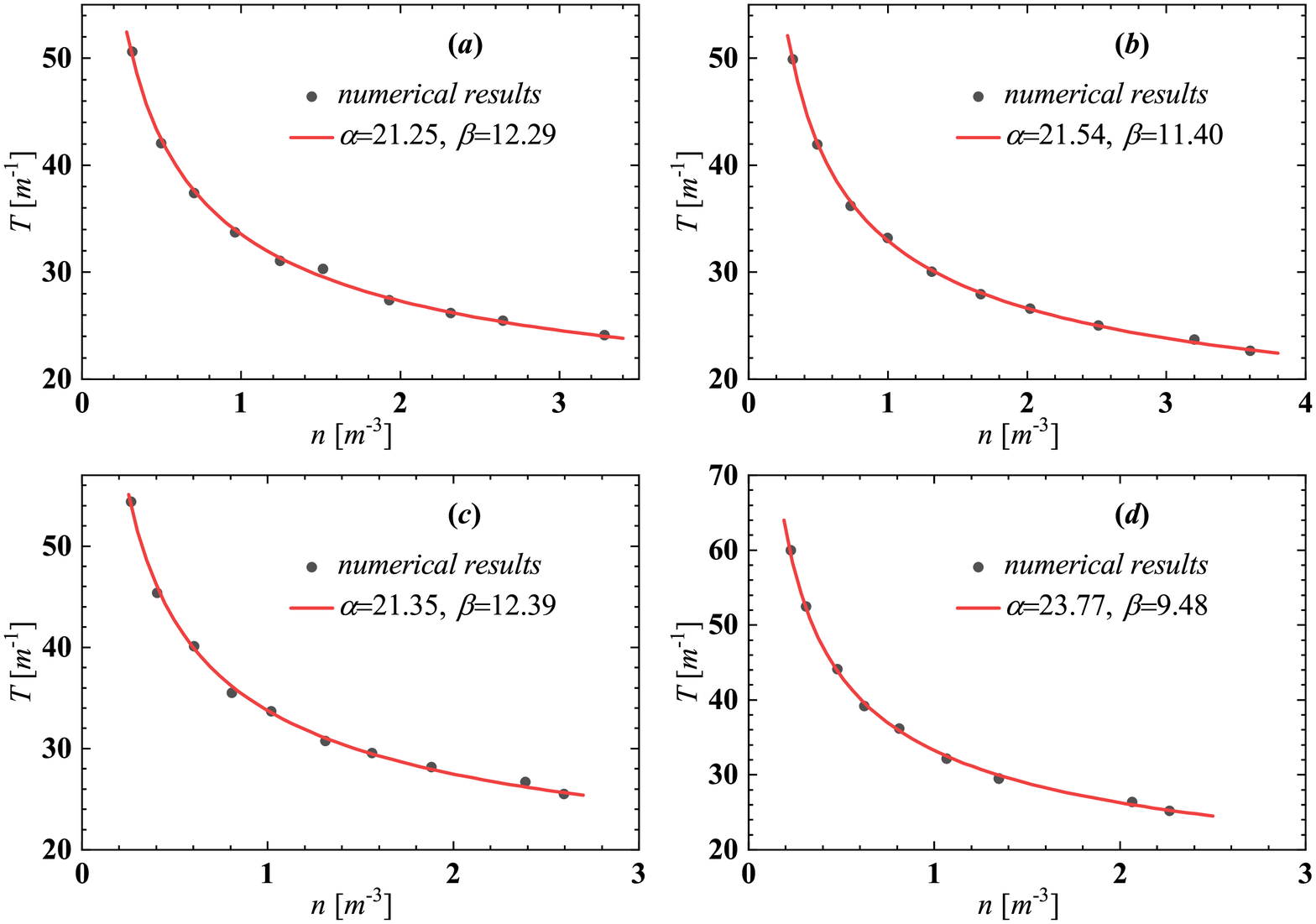}%
\caption{The oscillation period of electron-positron plasma varying with the particle number density. The $\omega$ and $\tau$ are fixed and the value of $E_{0}$ ranges from $5.0E_{\rm{cr}}$ to $14.0E_{\rm{cr}}$ with an interval of $1.0E_{\rm{cr}}$. Other external electric field parameters are (a) $\omega=1.5m$, $\tau=18.0/m$; (b) $\omega=1.5m$, $\tau=30.0/m$; (c) $\omega=1.7m$, $\tau=12.0/m$; (d) $\omega=2.0m$, $\tau=12.0/m$.
\label{fig:t1}}
\end{figure}

First, we keep $\omega$ and $\tau$ unchanged, and $E_0$ ranges from $5.0E_{\rm{cr}}$ to $14.0E_{\rm{cr}}$ with an interval of $1.0E_{\rm{cr}}$.
The oscillation period changing with the particle number density is shown in Fig. \ref{fig:t1}.
The black dots represent the numerical results where the oscillation period of internal electric field $T$ is estimated by Eq. (\ref{eq:eintfit}).
From the results, we can see that the particle number density always increases with the external field amplitude $E_0$ and the oscillation period $T$ decreases with the increase of the number density.
The red lines are the fitting curves corresponding to the fitting formula
\begin{eqnarray}\label{eq:tfit}
{T}(n) = \frac{\alpha }{\sqrt {n} } + \beta,
\end{eqnarray}
where $n$ is the particle number density, $\alpha$ and $\beta$ are fitting parameters.
Note that this formula is constructed based on the frequency of Langmuir oscillation \cite{Tonks1929} in plasma physics.
Although it is a little different from the ultrarelativistic formula given in Ref. \cite{Alkofer2001}, both of them have the same varying tendency.
In (a), (b), (c), and (d) of Fig. \ref{fig:t1}, the goodness of fit $R^2$ is about 0.998, 0.999, 0.998, and 0.999, respectively.
This shows that the formula above can fit the numerical results very well.
Furthermore, since the fitting parameters $\alpha$ and $\beta$ are almost independent of $E_0$, the oscillation period $T$ is not directly related to $E_0$.
Finally, we emphasize that the external electric field we considered here is a high-frequency strong field.
For other cases, the fitting formula (\ref{eq:tfit}) may fail.

\begin{figure}[!ht]
\centering
\includegraphics[width=0.45\textwidth]{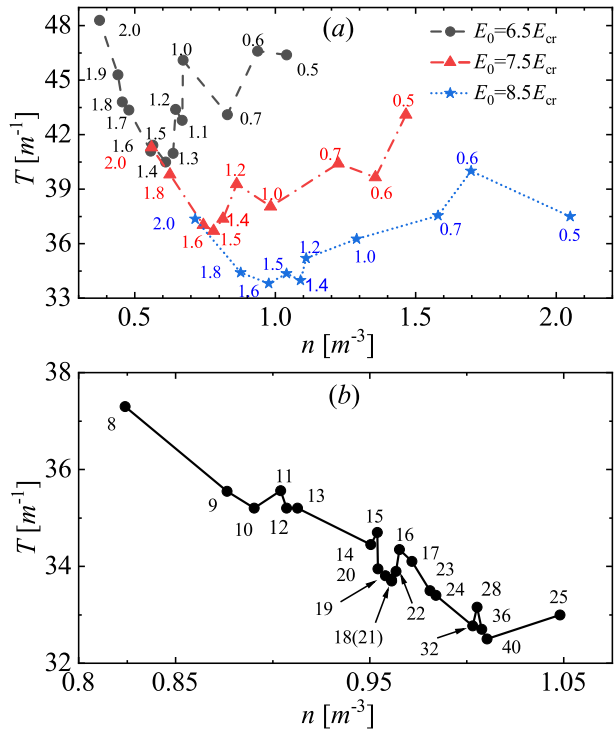}%
\caption{The oscillation period of electron-positron plasma varying with the particle number density. In (a), $E_0=6.5E_{\rm{cr}}$, $7.5E_{\rm{cr}}$, $8.5E_{\rm{cr}}$, and $\tau=12.0/m$ are fixed, and $\omega$ ranges from $0.5m$ to $2.0m$. The particle number density changes with $\omega$. In (b), $E_0=8.0E_{\rm{cr}}$ and $\omega=1.5m$ are constant, and the particle number density changes with $\tau$.
\label{fig:t2}}
\end{figure}

When $E_0$ and $\tau$ are kept unchanged and $\omega$ is changed, or keep $E_0$ and $\omega$ constant and change $\tau$, the above fitting formula will be invalid, see Fig. \ref{fig:t2}.
The relation between the oscillation period and the number density of created pairs is very complex and not monotonically decreasing.
In Fig. \ref{fig:t2}(a), we change the value of $\omega$ and mark each data point with the frequency of the external electric field.
It can be seen that the particle number density generally decreases with the increase of the frequency $\omega$.
In the high-frequency region, such as $\omega  \ge 1.6m$ for $E_0=6.5E_{\rm{cr}}$, $\omega  \ge 1.5m$ for $E_0=7.5E_{\rm{cr}}$, and $\omega  \ge 1.6m$ for $E_0=8.5E_{\rm{cr}}$, the oscillation period decreases with the increase of particle number density.
This behavior is somewhat similar to that in Fig. \ref{fig:t1}.
But when the frequency takes other values, the relation between them exhibits a very complex oscillations.
In particular, we find that in the low frequency region (such as $\omega=0.6m$) the fitting formula (\ref{eq:tfit}) still fails even if the field frequency and the pulse duration are fixed and only the field amplitude changes, because the number density does not decrease monotonically with the field amplitude.
In Fig. \ref{fig:t2}(b), we change the value of $\tau$ and mark each data point with the value of $\tau$.
It can be seen that the relationship between the particle number density and the pulse duration $\tau$ is not monotonic.
As the pulse duration becomes large, the number density of created pairs may not increase.
Thus, the fitting formula (\ref{eq:tfit}) also does not work here.
In addition, from Fig. \ref{fig:t2}(a), we can see that for $E_0=6.5E_{\rm{cr}}$ the number density of created particles at $\omega=1.1m$ almost equals that at $\omega=1.0m$, but their oscillation periods have a big difference. That is to say, the particle number density for different field frequencies can be equal, see Fig. \ref{fig:number2}, but the same number density corresponds to different oscillation periods, which shows that the oscillation periods directly depend on not only the particle number density but also the field frequency.
Similarly, from Fig. \ref{fig:t2}(b), we can find that the oscillation periods also directly depend on the pulse duration, because the different number density corresponding to different pulse durations gives the same oscillation period, see the data points marked with the pulse duration $12$ and $13$.

\begin{figure}[!ht]
\centering
\includegraphics[width=0.45\textwidth]{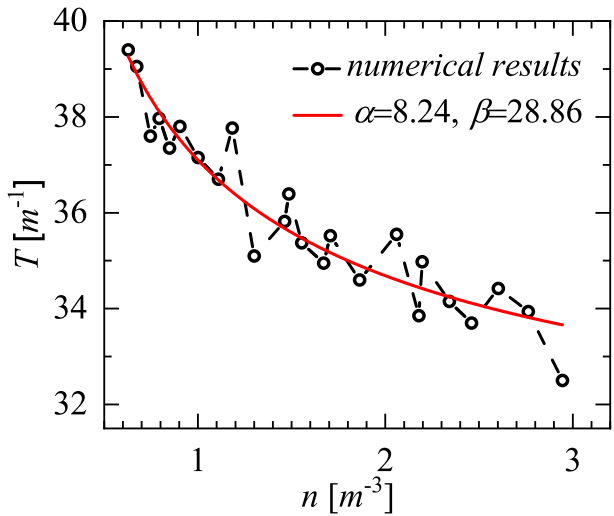}%
\caption{The oscillation period of electron-positron plasma changing with the particle number density. Here $E_0=7.0E_{\rm{cr}}$, $\omega \tau = 20.0$ is kept to ensure that the number of cycles in the external electric field is constant. The value of $\omega$ ranges from $0.4m$ to $1.65m$ and $\tau$ changes accordingly.
\label{fig:t3}}
\end{figure}

We further explore the relation between the oscillation period and the particle number density for fixing $E_0=7.0E_{\rm{cr}}$ and keeping $\omega \tau = 20.0$, which ensures that the number of cycles in the external electric field is constant, see Fig. \ref{fig:t3}.
To separate the number density the value of $\omega$ ranges from $0.4m$ to $1.65m$ with a variable interval.
From the figure, we find that with the frequency increasing (the pulse duration decreasing correspondingly), the particle number density always decreases.
When the value of $\omega$ is relatively small, the particle number density changes faster with $\omega$.
For example, when $\omega$ changes from $0.41m$ to $0.4m$, the particle number density changes $0.1817m^{-3}$, while when $\omega$ changes from $1.5m$ to $1.4m$, the number density only changes $0.0725m^{-3}$.
This result is also reflected in Fig. \ref{fig:distribution}(b).
Besides this, the oscillation period tends to decrease with the increase of the number density.
Therefore, we can try to fit it with Eq. (\ref{eq:tfit}).
The fitted curve is represented by the solid red line in Fig. \ref{fig:t3}, and the goodness of fit $R^2$ is about $0.89$.
This suggests that considering the product of $\omega$ and $\tau$ as a whole may be more helpful to explore the relation between the oscillation period and the number density of produced pairs.

It is noted that the approximate expression of the internal electric field Eq. (\ref{eq:eintfit}) is under the premise that the external field is zero, but the plasma oscillation period is still directly related to the external field parameters $E_0$, $\omega$ and $\tau$, which also embodies the non-Markov effect in the electron-positron pair production.

\section{CONCLUSION AND DISCUSSION}
\label{sec:four}

In summary, the backreaction effect and plasma oscillation in the pair production for a high-frequency electric field are investigated by the QVE.

First, the parameter region of the external electric field where the backreaction effect cannot be ignored is explored by calculating and analyzing the difference degree of pair number density with and without the backreaction.
It is found that the backreaction effect can be neglected in the pair production for a high-frequency but weak electric field.
In other words, the backreaction effect can be ignored in the pair production for a subcritical external field with frequency chirp.
This result is beyond what we previously thought, because the great improvement of particle number density by the high-frequency photon absorption process is expected to make the backreaction effect become obvious.
The reason are as follows. When the external electric field strength is smaller than the critical field strength, no matter how large the field frequency is, the backreaction effect can be neglected.
Thus, to study the backreaction effect, the external electric field strength should be larger than the critical one.
However, for a high-frequency and strong electric field, the pair production is dominated by tunneling pair production and the multiphoton absorption mechanism at the same time, which will suppress the pair production because of the competitive relation between these two mechanisms.

The influence of backreaction on the momentum spectrum is also considered.
It is found that the change of the momentum spectrum is very complex and sensitive to the parameters of the external electric field.

Finally, the relationship between the plasma oscillation period and the number density of produced particle pairs is studied.
When the frequency and duration of the external electric field are kept constant and only the field strength is changed, the relation between the oscillation period and the particle number density is obvious and can be fitted by a simple formula constructed based on the Langmiur oscillation frequency.
However, when the field frequency or the pulse duration changes, the relationship between them is very complex.
One way to ensure that the oscillation frequency is regular is to keep the number of cycles in the external electric field unchanged when changing the field frequency.
Furthermore, we find that the plasma oscillation period not only directly depend on the final number density of created particles, but also depend on the external field parameters, such as the field strength, the frequency, and the pulse duration, which is different from the common result in plasma physics. The reason for this is that the external electric field has left an imprint on the internal field by the non-Markov effect in pair production.

Our results clarify the question whether it is reasonable to study the pair production for a subcritical electric field with frequency chirp, and deepen the understanding of the influence factors of the plasma oscillation period.

\begin{acknowledgments}
The work is supported by the National Natural Science Foundation of China (NSFC) under Grants No. 11974419 and No. 11705278, and by the Fundamental Research Funds for the Central Universities (20226943).
\end{acknowledgments}


\begin{thebibliography}{99}

\bibitem{Dirac1928}
 P. A. M. Dirac, Proc. Roy. Soc. Lond. A \textbf{117}, 610 (1928).

\bibitem{Sauter1931}
 F. Sauter, Z. Phys. \textbf{69}, 742 (1931).

\bibitem{Schwinger1951}
 J. Schwinger, On Gauge Invariance and Vacuum Polarization, Phys. Rev. \textbf{82} 664 (1951).

\bibitem{Xie2017}
B. S. Xie, Z. L. Li, and S. Tang, Electron-positron pair production in ultrastrong laser fields, Matter and Radiation at Extremes \textbf{2}, 225 (2017).

\bibitem{Fedotov2022}
A. Fedotov, A. Ilderton, F. Karbstein, B. King, D. Seipt, H. Taya, and G. Torgrimsson, Advances in QED with intense background fields, arXiv:2203.00019 [hep-ph]

\bibitem{Tanaka2019}
K. A. Tanaka, K. M. Spohr, D. L. Balabanski, S. Balascuta, L. Capponi, M. O. Cernaianu, M. Cuciuc, A. Cucoanes, I. Dancus, A. Dhal, B. Diaconescu, D. Doria, P. Ghenuche, D. G. Ghita, S. Kisyov, V. Nastasa, J. F. Ong, F. Rotaru, D. Sangwan, P.-A. S\"{o}derstr\"{o}m, D. Stutman, G. Suliman, O. Tesileanu, L. Tudor, N. Tsoneva, C. A. Ur, D. Ursescu, and N. V. Zamfir, Current status and highlights of the ELI-NP research program,  Matter and Radiation at Extremes \textbf{5}, 024402(2019).

\bibitem{Tiwari2019}
G. Tiwari, E. Gaul, M. Martinez, G. Dyer, J. Gordon, M. Spinks, T. Toncian, B. Bowers, X. Jiao, R. Kupfer, L. Lisi, E. McCary, R. Roycroft, A. Yandow, G. D. Glenn, M. Donovan, T. Ditmire, and B. M. Hegelich, Beam distortion effects upon focusing an ultrashort petawatt laser pulse to greater than $10^{22}\rm{W}/\rm{cm}^2$, Opt. Lett. \textbf{44}, 2764(2019).

\bibitem{Yoon2021}
J. W. Yoon, Y. G. Kim, I. W. Choi, J. H. Sung, H. W. Lee, S. K. Lee, C. H. Nam, Realization of laser intensity over $10^{23}\rm{W}/\rm{cm}^2$, Optica \textbf{8}, 630(2021).

\bibitem{Ringwald2001}
 A. Ringwald, Pair production from vacuum at the focus of an X-ray free electron laser, Phys. Lett. B \textbf{510}, 107 (2001).

\bibitem{Schutzhold2008}
 R. Sch\"{u}tzhold, H. Gies, and G. Dunne, Dynamically Assisted Schwinger Mechanism, Phys. Rev. Lett. \textbf{101}, 130404 (2008).

\bibitem{Nuriman2012}
 A. Nuriman, B. S. Xie, Z. L. Li, and D. Sayipjamal, Enhanced electron-positron pair creation by dynamically assisted combinational fields, Phys. Lett. B \textbf{717}, 465 (2012).

\bibitem{Fey2012}
C. Fey and R. Sch\"{u}tzhold, Momentum dependence in the dynamically assisted Sauter-Schwinger effect, Phys. Rev. D \textbf{85}, 025004 (2012).

\bibitem{Kohlfurst2013}
C. Kohlf\"{u}rst, M. Mitter, G. von Winckel, F. Hebenstreit, and R. Alkofer, Optimizing the pulse shape for Schwinger pair production, Phys. Rev. D \textbf{88}, 045028 (2013).

\bibitem{Li2014}
Z. L. Li, D. Lu, B. S. Xie, L. B. Fu, J. Liu, and B. F. Shen, Enhanced pair production in strong fields by multiple-slit interference effect with dynamically assisted Schwinger mechanism, Phys. Rev. D \textbf{89}, 093011 (2014).

\bibitem{Jiang2013}
M. Jiang, B. S. Xie, H. B. Sang, and Z. L. Li, Enhanced electron-positron pair creation by the frequency chirped laser pulse, Chin. Phys. B \textbf{22}, 100307 (2013).

\bibitem{Abdukerim2017}
N. Abdukerim, Z. L. Li, and B. S. Xie, Enhanced electron-positron pair production by frequency chirping in one- and two-color laser pulse fields, Chin. Phys. B \textbf{26}, 020301 (2017).

\bibitem{Gong2020}
C. Gong, Z. L. Li, B. S. Xie, and Y. J. Li, Electron-positron pair production in frequency modulated laser fields, Phys. Rev. D \textbf{101}, 016008 (2020).

\bibitem{Olugh2019}
O. Olugh, Z. L. Li, B. S. Xie, and R. Alkofer, Pair production in differently polarized electric fields with frequency chirps, Phys. Rev. D \textbf{99}, 036003 (2019).

\bibitem{Wang2021}
K. Wang, X. Hu, S. Dulat, and B. S. Xie, Effect of symmetrical frequency chirp on pair production, Chin. Phys. B \textbf{30}, 060204 (2021).

\bibitem{Li2021}
L. J. Li, M. Mohamedsedik, and B. S. Xie, Enhanced dynamically assisted pair production in spatial inhomogeneous electric fields with the frequency chirping, Phys. Rev. D \textbf{104}, 036015 (2021).

\bibitem{Mohamedsedik2021}
M. Mohamedsedik, L. J. Li, and B. S. Xie, Schwinger pair production in inhomogeneous electric fields with symmetrical frequency chirp, Phys. Rev. D \textbf{104}, 016009 (2021).

\bibitem{Alkofer2001}
R. Alkofer, M. B. Hecht, C. D. Roberts, S. M. Schmidt, and D. V. Vinnik, Pair Creation and an X-Ray Free Electron Laser, Phys. Rev. Lett. \textbf{87}, 193902 (2001).

\bibitem{Bloch1999}
J. C. R. Bloch, V. A. Mizerny, A. V. Prozorkevich, C. D. Roberts, S. M. Schmidt, S. A. Smolyansky, and D. V. Vinnik, Pair creation: Back reactions and damping, Phys. Rev. D \textbf{60}, 116011 (1999).

\bibitem{Roberts2002}
C. D. Roberts, S. M. Schmidt, and D. V. Vinnik, Quantum Effects with an X-Ray Free-Electron Laser, Phys. Rev. Lett. \textbf{89}, 153901 (2002).

\bibitem{Pla2021}
S. Pla, I. M. Newsome, R. S. Link, P. R. Anderson and J. Navarro-Salas, Pair production due to an electric field in $1+1$ dimensions and the validity of the semiclassical approximation, Phys. Rev. D \textbf{103}, 105003(2021).

\bibitem{Kohlfurst2014}
C. Kohlf\"{u}rst, H. Gies, and R. Alkofer, Effective Mass Signatures in Multiphoton Pair Production, Phys. Rev. Lett. \textbf{112}, 050402 (2014).

\bibitem{Keldysh}
L. V. Keldysh, Ionization in the field of a strong electromagnetic wave, Sov. Phys. JETP \textbf{20}, 1307 (1965).

\bibitem{Benedetti2011}
A. Benedetti, W. B. Han, R. Ruffini, and G. V. Vereshchagin, On the frequency of oscillations in the pair plasma generated by a strong electric field, Phys. Lett. B \textbf{698}, 75 (2011).

\bibitem{Tonks1929}
L. Tonks and I. Langmuir, Oscillations in Ionized Gases, Phys. Rev. \textbf{33}, 195 (1929).


\end{thebibliography}
\end{document}